\documentclass[bibyear]{aa}

\usepackage{amsmath}
\usepackage{graphicx}
\usepackage{txfonts}
\usepackage{setspace}

\begin{document}

\title{Sublimation Origin of Active Asteroid P/2018 P3}

\author{Yoonyoung Kim\inst{1}
	\and Jessica Agarwal\inst{1,2}
	\and David Jewitt\inst{3}
	\and Max Mutchler\inst{4}
	\and Stephen Larson\inst{5}
	\and Harold Weaver\inst{6}
	\and Michael Mommert\inst{7}}

\institute{Institute for Geophysics and Extraterrestrial Physics, TU Braunschweig, 38106 Braunschweig, Germany\\
\email{yoonyoung.kim@tu-bs.de}
\and Max Planck Institute for Solar System Research, 37077 G\"ottingen, Germany
\and Department of Earth, Planetary and Space Sciences,
UCLA, Los Angeles, CA 90095-1567, USA
\and Space Telescope Science Institute, Baltimore, MD 21218, USA
\and Lunar and Planetary Laboratory, University of Arizona, Tucson, AZ 85721-0092, USA
\and The Johns Hopkins University Applied Physics Laboratory,  Laurel, Maryland 20723, USA
\and University of St. Gallen, Institute of Computer Science,  9000 St. Gallen, Switzerland}

\abstract
{Active asteroids show (typically transient) cometary activity, driven by a range of processes.  A sub-set, sometimes called main-belt comets, may be driven by sublimation and so could be useful for tracing the present-day distribution of asteroid ice.
Object P/2018 P3 has a Tisserand parameter 3.096 but a high eccentricity 0.415, placing it within the dynamical boundary between asteroids and comets.}
{We aim to determine the cause of activity (sublimation or something else) and assess the dynamical stability of P3, in order to better constrain the intrinsic ice content in the main belt.
}
{We obtained Hubble Space Telescope images of P3 at the highest angular resolution. We compared the observations with a Monte Carlo model of dust dynamics. We identified and analyzed archival CFHT (2013) and NEOWISE (2018) data.
In addition, we numerically integrated the orbits of P3 clones  for 100 Myr.}
{P3 has been recurrently active near two successive perihelia (at 1.76 AU), indicative of a sublimation origin. The absence of 4.6~$\mu$m band excess indicates zero or negligible CO or CO$_2$ gas production from P3.
The properties of the ejected dust are remarkably consistent with those found in other main-belt comets (continuous emission of $\sim$0.05--5 mm particles at 0.3--3 m s$^{-1}$ speeds), with mass-loss rates of $\gtrsim$2 kg s$^{-1}$. The orbit of P3 is unstable on timescales $\sim$10 Myr.}
{We speculate that P3 has recently arrived from a more stable source (either the Kuiper Belt or elsewhere in the main belt) and has been physically aged at its current location, finally becoming indistinguishable from a weakly sublimating asteroid in terms of its dust properties. Whatever the source of P3, given the dynamical instability of its current orbit, P3 should not be used to trace the native distribution of asteroid ice.}


\titlerunning{Active Asteroid P/2018 P3}
\authorrunning{Y. Kim et al.}
\maketitle

\section{INTRODUCTION}

Active asteroids are small solar system bodies that combine asteroid-like orbits and comet-like activity (Jewitt et al. 2015). By definition, active asteroids have Tisserand parameter values of $T_J > 3.08$ (i.e. dynamically decoupled from Jupiter and excludes Encke-type comets), while Kuiper Belt comets and Oort cloud comets have $T_J < 3$.
Mass loss mechanisms identified to date include sublimation, impacts, rotational breakup, and combinations of these processes.  A subset of the active asteroids called main-belt comets (MBCs; Hsieh \& Jewitt 2006) exhibits recurrent mass loss near perihelion, indicating sublimation-driven activity.
Proper identification of MBCs is essential to improve our understanding of the distribution and abundance of ice in the main belt.

Active asteroid P/2018 P3 (PANSTARRS, hereafter ``P3'') was discovered in an active state on UT 2018 August 08 (Weryk et al.~2018), two months before perihelion on UT 2018 October 09.
Its orbital semimajor axis, eccentricity and inclination are 3.007 AU, 0.415 and 8.90\degr, respectively, leading to an asteroid-like Tisserand parameter, $T_J$ = 3.096.
While most of the currently known active asteroids have relatively low eccentricities and are assumed to be native to the main belt, the orbital eccentricity of P3 is comparatively high, which favors excavation and sublimation of sub-surface ice through higher impact velocities and perihelion temperatures (Kim et al. 2018) but also increases the probability for an origin in the Kuiper Belt (Hsieh \& Haghighipour 2016).

In this paper we report time-resolved observations from the Hubble Space Telescope (HST) taken to investigate P3 at high spatial resolution.
We also aim to determine the cause of the activity and assess the dynamical stability of P3, ultimately to better constrain the native population of main-belt ice.

\begin{table*}
\caption{Observing Geometry 
\label{geometry}}
\centering
\begin{tabular}{lcccrccccr}
\hline\hline
UT Date and Time & DOY\tablefootmark{a}   & $\Delta T_p$\tablefootmark{b} & $\nu$\tablefootmark{c} & $r_H$\tablefootmark{d}  & $\Delta$\tablefootmark{e} & $\alpha$\tablefootmark{f}   & $\theta_{\odot}$\tablefootmark{g} &   $\theta_{-v}$\tablefootmark{h}  & $\delta_{\oplus}$\tablefootmark{i}\\
\hline
2018 Sep 28 18:19 - 18:56 & 271 & -11 & 354.6  	&  1.758 	& 0.786 	& 11.9 	& 11.9 	& 243.2 	& 9.1 \\
2018 Nov 14 20:07 - 20:46 & 318 & 36 & 18.1  	&  1.782 	& 1.035 	& 27.5 	& 60.2 	& 242.6 	& 1.0 \\
2018 Dec 28 20:54 - 21:32 &   362 & 80 &  38.6 & 1.877 & 1.503 & 31.4 & 66.8 & 239.7 & -3.5 \\
\hline
\end{tabular}
\tablefoot{
\tablefoottext{a}{Day of Year, UT 2018 January 01 = 1.}
\tablefoottext{b}{Number of days from perihelion (UT 2018-Oct-09 = DOY 282).}
\tablefoottext{c}{True anomaly, in degrees.}
\tablefoottext{d}{Heliocentric distance, in AU.}
\tablefoottext{e}{Geocentric distance, in AU.}
\tablefoottext{f}{Phase angle, in degrees.}
\tablefoottext{g}{Position angle of the projected anti-Solar direction, in degrees.}
\tablefoottext{h}{Position angle of the projected negative heliocentric velocity vector, in degrees.}
\tablefoottext{i}{Angle of Earth above the orbital plane, in degrees.}
}
\end{table*}

\begin{figure*}
\centering
\includegraphics[width=0.9\textwidth]{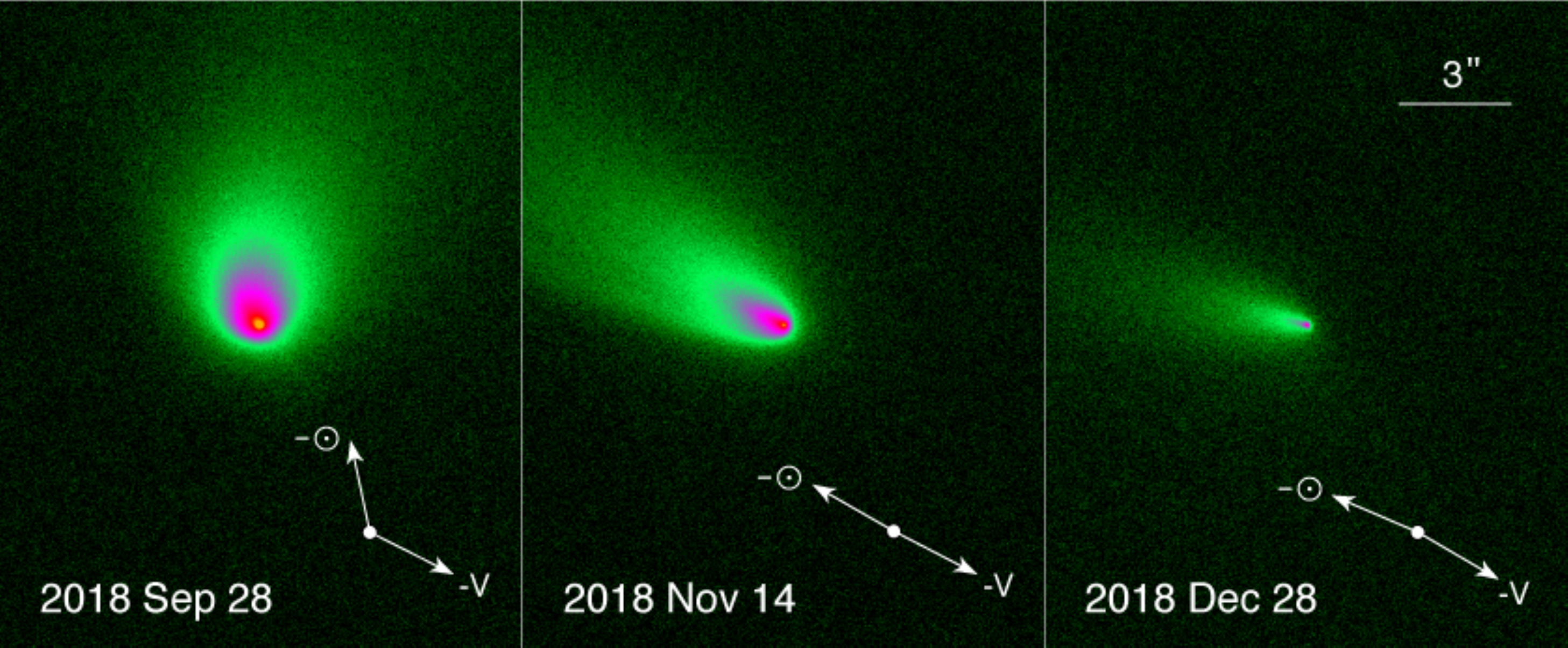}
\caption{Composite HST images of P3 marked with UT dates of observation.
A scale bar, the projected anti-solar direction ($-\odot$) and the negative heliocentric velocity vector ($-V$) are indicated. Celestial north points up and east points to the left.
\label{images}}
\end{figure*}

\section{OBSERVATIONS}

Observations with the HST were taken under the Director's Discretionary Time allocation GO 15623. 
We used the UVIS channel of the WFC3 camera with the broadband F606W filter (effective wavelength $\sim$6000\AA, FWHM $\sim$2300\AA).
In each HST orbit we obtained eight exposures of 230 s duration using a 1k subarray of WFC3 (40\arcsec$\times$40\arcsec~field), with a 2$\times$2 subsampling dither-pattern.  The image scale was initially 0.04\arcsec~pixel$^{-1}$ and we re-sampled the images to have a scale of 0.025\arcsec~pixel$^{-1}$.

The remarkably small geocentric distance (0.786 AU) on UT 2018 September 28 provides superb spatial resolution of 17 km per re-sampled  WFC3 pixel, giving the opportunity to study the inner coma of an active asteroid near perihelion in great detail.
The observations on UT 2018 November 14 were scheduled to coincide with the passage of the Earth through the orbital plane (the out-of-plane angle was 1\degr) to measure the distribution of dust perpendicular to the orbit plane.
A journal of observations is given in Table \ref{geometry}.

\begin{figure}
\centering
\includegraphics[width=1.0\columnwidth]{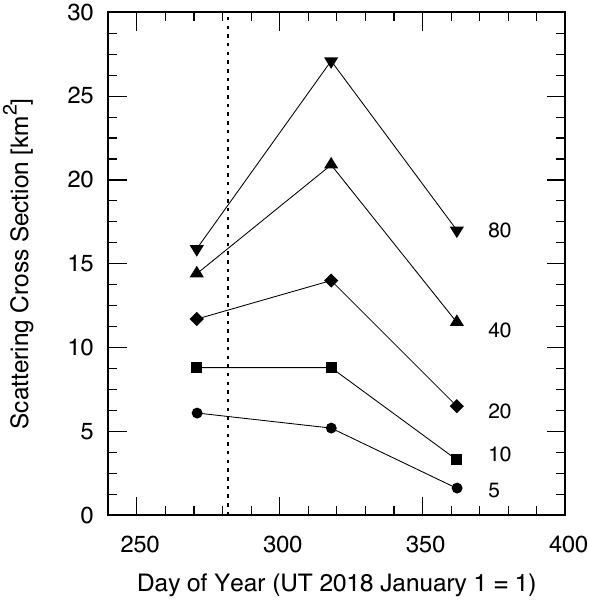}
\caption{Scattering cross-section as a function of time, expressed as Day of Year (DOY=1 on UT 2018 January 1).
The radii of the apertures (in units of 10$^2$ km) are marked.   The dotted line indicates the date of perihelion, UT 2018 October 09. \label{ce}}
\end{figure}

\begin{table*}
\caption{Photometry with Fixed Linear Radius Apertures
\label{phot}}
\centering
\begin{tabular}{lccccccc}
\hline\hline
UT Date    & Quantity\tablefootmark{a} & 500 km    & 1000 km & 2000 km & 4000 km & 8000 km \\
\hline
Sep 28 	&	$V$ 			& 18.16$\pm$0.01	 	& 17.76$\pm$0.01	 	& 17.45$\pm$0.03	 		& 17.22$\pm$0.03	 		& 17.11$\pm$0.03	 		 \\
Sep 28   & $H$ 			& 16.98$\pm$0.01		& 16.58$\pm$0.01	 	& 16.27$\pm$0.03	 		& 16.04$\pm$0.03	 	& 15.93$\pm$0.03	 \\
Sep 28 	& $C_e$ 			& 6.1$\pm$0.2 		& 8.8$\pm$0.5	& 11.7$\pm$0.6			& 14.4$\pm$0.6		& 15.9$\pm$0.6	 \\\\
Nov 14 	&	$V$ 			& 19.57$\pm$0.01	& 19.01$\pm$0.01	 	& 18.50$\pm$0.03	 		& 18.06$\pm$0.06	 		& 17.78$\pm$0.06	 		 \\
Nov 14   & $H$ 			& 17.14$\pm$0.01		& 16.58$\pm$0.01	 	& 16.07$\pm$0.03	 		& 15.63$\pm$0.06	 	& 15.35$\pm$0.06	 \\
Nov 14 	& $C_e$ 			& 5.2$\pm$0.2 		& 8.8$\pm$0.4		& 14.0$\pm$0.5			& 20.9$\pm$0.4		& 27.1$\pm$0.4	 \\\\
Dec 28 	&	$V$ 			& 21.90$\pm$0.01	 	& 21.16$\pm$0.01	 	& 20.42$\pm$0.03	 		& 19.79$\pm$0.06	 		& 19.37$\pm$0.06	 		 \\
Dec 28   & $H$ 			& 18.39$\pm$0.01		& 17.65$\pm$0.01	 	& 16.91$\pm$0.03	 		& 16.28$\pm$0.06	 	& 15.86$\pm$0.06	 \\
Dec 28 	& $C_e$ 			& 1.6$\pm$0.3 		& 3.3$\pm$0.6		& 6.5$\pm$0.8			& 11.5$\pm$0.7		& 17.0$\pm$0.7	 \\   
\hline
\end{tabular}
\tablefoot{
\tablefoottext{a}{$V$ = total apparent V magnitude, $H$ = total absolute V magnitude, $C_e$ = effective scattering cross-section in km$^2$ computed with $p_V$ = 0.04.}
}
\end{table*}

\begin{figure*}
\centering
\includegraphics[width=1.0\textwidth]{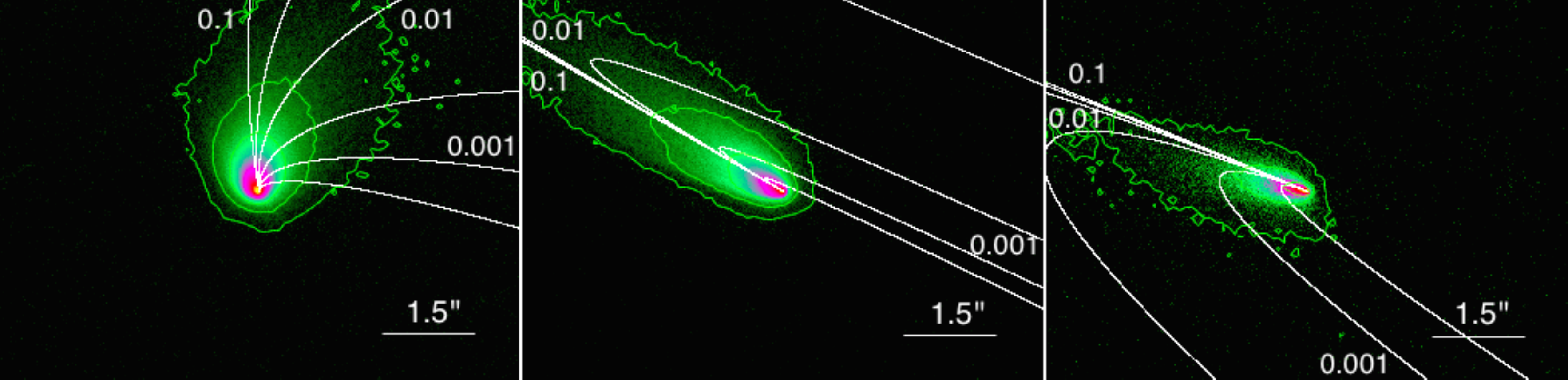}
\caption{Syndynes, showing the trajectories of particles with $\beta=$ 0.1, 0.03, 0.01, 0.003, 0.001, and 0.0003, for the three images shown in Figure \ref{images}. Celestial north points up and east points to the left.
\label{syn}}
\end{figure*}

\section{RESULTS}

\subsection{Photometry}

We obtained photometry from each composite image (Figure \ref{images}) using a set of five circular apertures having fixed linear radii from 500 to 8,000~km, projected to the distance of P3.
The sky background was determined within a concentric annulus with inner and outer radii of 15\arcsec~and 20\arcsec, respectively.
Flux calibration was performed using the online WFC3 Exposure Time calculations for a G2V source in the F606W filter.
We converted the total apparent magnitudes, $V$, to total absolute magnitudes, $H$, using

\begin{equation}
H = V - 5\log_{10}(r_H \Delta) - f(\alpha).
\label{absolute}
\end{equation}

\noindent where $r_H$ and $\Delta$ are the heliocentric and geocentric distances, respectively. 
The phase function $f(\alpha)$ at solar phase angle $\alpha$ was not measured for P3. We used a linear phase function $f(\alpha) = 0.04\alpha$ (Meech \& Jewitt 1987).
The total absolute magnitude is related to the effective scattering cross-section, $C_e$ [km$^2$], by
\smallskip

\begin{equation}
C_e = \frac{1.5\times 10^6}{p_V} 10^{-0.4 H}
\end{equation}

\noindent where $p_V$ = 0.04 is the assumed geometric albedo (Fern{\'a}ndez et al. 2013).
For each date and aperture radius, $V$, $H$, and $C_e$ are listed in Table \ref{phot}.

In Figure \ref{ce}, the scattering cross-section within the 8,000~km radius aperture brightened by $\Delta C_e$~=~11 km$^2$ from 16 km$^2$ in September (10 days before perihelion) to 27 km$^2$ in November (1 month after perihelion), then dropped back to 17 km$^2$ at the end of December.
The scattering cross-section within the central aperture decreased continuously from September to December, reaching $C_e$ = 1.6 km$^2$ on UT 2018 December 28.
A crude upper limit to the nucleus radius (using photometry from the central aperture) is given by $r_n = (C_e/\pi)^{1/2}$ = 0.71 km, assuming $p_V$ = 0.04.

\subsection{Morphology}
\label{morphology}

For a density of 1000 kg m$^{-3}$, the Hill radius of an object having a radius of $r_n$ = 700~m and the orbit of P3 (heliocentric distance $r_H$ = 1.75 AU) is $\sim$200 km. 
We attempted to search for a potential binary companion and large unbound fragments since our high-resolution images should have good resolution of the Hill sphere (corresponding to $\sim$11 WFC3 pixels).
A wide binary system (Agarwal et al. 2020), and gravitationally unbound fragments resulting from a (near-)catastrophic collision (Kim et al. 2017) or spin-up
(Drahus et al. 2015) have been found in other active asteroids, providing key constraints determining the cause of activity.

Observations on all dates show a comet-like coma (Figure \ref{images}), with no obvious sign of fragments or other structures in the coma.
To enhance the inner coma structure, we subtracted a 1/$\rho$ gradient and azimuthal average using the Cometary Coma Image Enhancement Facility (Samarasinha et al. 2013).
The results revealed an excess jet-like structure, with a jet axis rotating counter-clockwise that closely follows the changing antisolar direction (Figure \ref{images}).
This is consistent with recently-released particles that are small enough to be strongly accelerated by solar radiation pressure, indicating continuous dust emission. In contrast, we observe in the enhanced images no evidence for binarity or companions/fragments.
To set a limit to the size of unseen secondary objects, we used the on-line Exposure Time Calculator to find that, in a 8 $\times$ 230 s integration, signal-to-noise ratio SNR = 10 is reached at magnitude $V$ = 26.5.  The corresponding limiting absolute magnitude is $H >$ 25.3 and the upper limit to the radius is $r_e <$  26 m (geometric albedo $p_V$ = 0.04 assumed).  The limiting magnitude will be poorer in regions of the image where scattered light from dust elevates the background brightness, but we have not attempted to quantify this effect.

The motion of dust particles in interplanetary space is controlled by $\beta$, the ratio of radiation pressure acceleration to solar gravity.  $\beta$ is a function of particle size, approximately given by $\beta \sim a^{-1}$, where $a$ is the particle radius expressed in microns.  
Figure \ref{syn}  shows syndyne trajectories, which are the loci of particles of a given $\beta$ released from the nucleus with zero ejection velocity at different times (Finson \& Probstein 1968).
The directions of the tail are best matched by syndynes with $\beta \sim$ 0.003 to 0.03, corresponding to particle radii $a \sim$ 30 to 300 $\mu$m.  We take $a \sim$ 100 $\mu$m as the nominal grain size in P3.

\begin{figure}
\centering
\includegraphics[width=0.95\columnwidth]{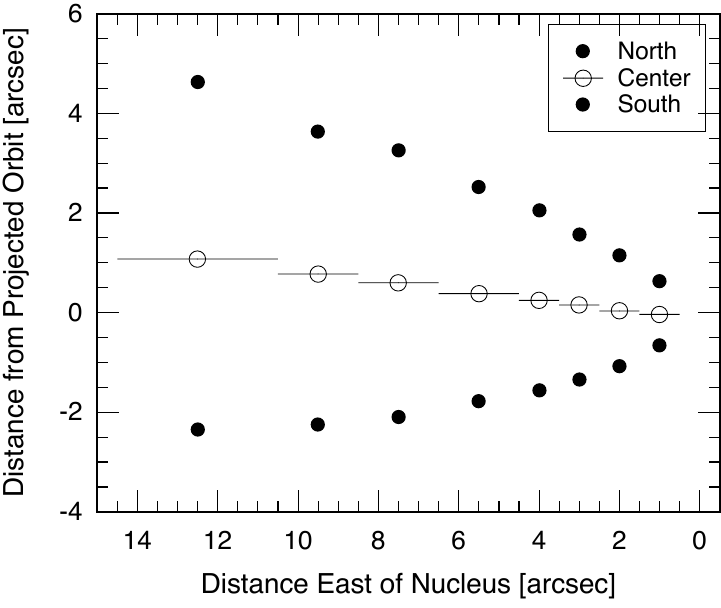}
\caption{Distance from projected orbit as a function of distance east of the nucleus, observed on UT 2018 November 14. The positions of peak and half-peak brightness are measured separately to the north and south of the tail. Horizontal bars show the width of each segment used to average the signal.
\label{profile}}
\end{figure}

\subsection{Dust Profiles}
\label{perpendicular}

Observations on UT 2018 November 14  were taken only +1$\degr$ from the projected orbital plane of P3, and provides a constraint on the dust ejection velocity component perpendicular to the orbital plane.  
For this purpose, we computed surface brightness profiles cut perpendicular to the tail direction in the image plane, in  1--4\arcsec~wide segments, where the tail gradually widened with distance from the nucleus.

Figure \ref{profile} shows the positions of peak and half-peak brightness measured separately to the north and south of the tail. 
Because of the projection effect that occurs when viewing the dust sheet from  above the plane (even though by only +1$\degr$), the dust extends slightly further north than south of the tail axis (Figure \ref{syn}). Nevertheless, the profile (Figure \ref{profile}) appeared symmetrical from the peak dust brightness, showing that the ejection is almost symmetric.
We take the FWHM of a series of vertical profiles as the measure of the out-of-plane width.

The width of the tail, $w_T$, is related to the distance from the nucleus, $\ell_T$ [m], by

\begin{equation}
V_{\perp} = \left(\frac{\beta g_{\odot}}{8 \ell_T} \right)^{1/2} w_T
\label{width}
\end{equation}

\noindent  where $V_{\perp}$ is the ejection velocity perpendicular to the orbit plane and $g_{\odot} \sim 0.002$ m s$^{-2}$ is the local solar gravitational acceleration at $r_H$ = 1.78 AU.  
For simplicity, we assume that $\ell_T$ is proportional to the projected angular distance from the nucleus, $\theta$.
We fitted Equation (\ref{width}) to the FWHM in Figure \ref{profile}, finding $V_{\perp}$ = (20$\pm$2) m s$^{-1}$ on the tail for $\beta = 1$ particles.
Within the uncertainties, we take $V_{\perp} \sim 20\sqrt{\beta}$ m s$^{-1}$  as the dust ejection velocity.

\section{DISCUSSION}

\begin{figure}
\centering
\includegraphics[width=0.95\columnwidth]{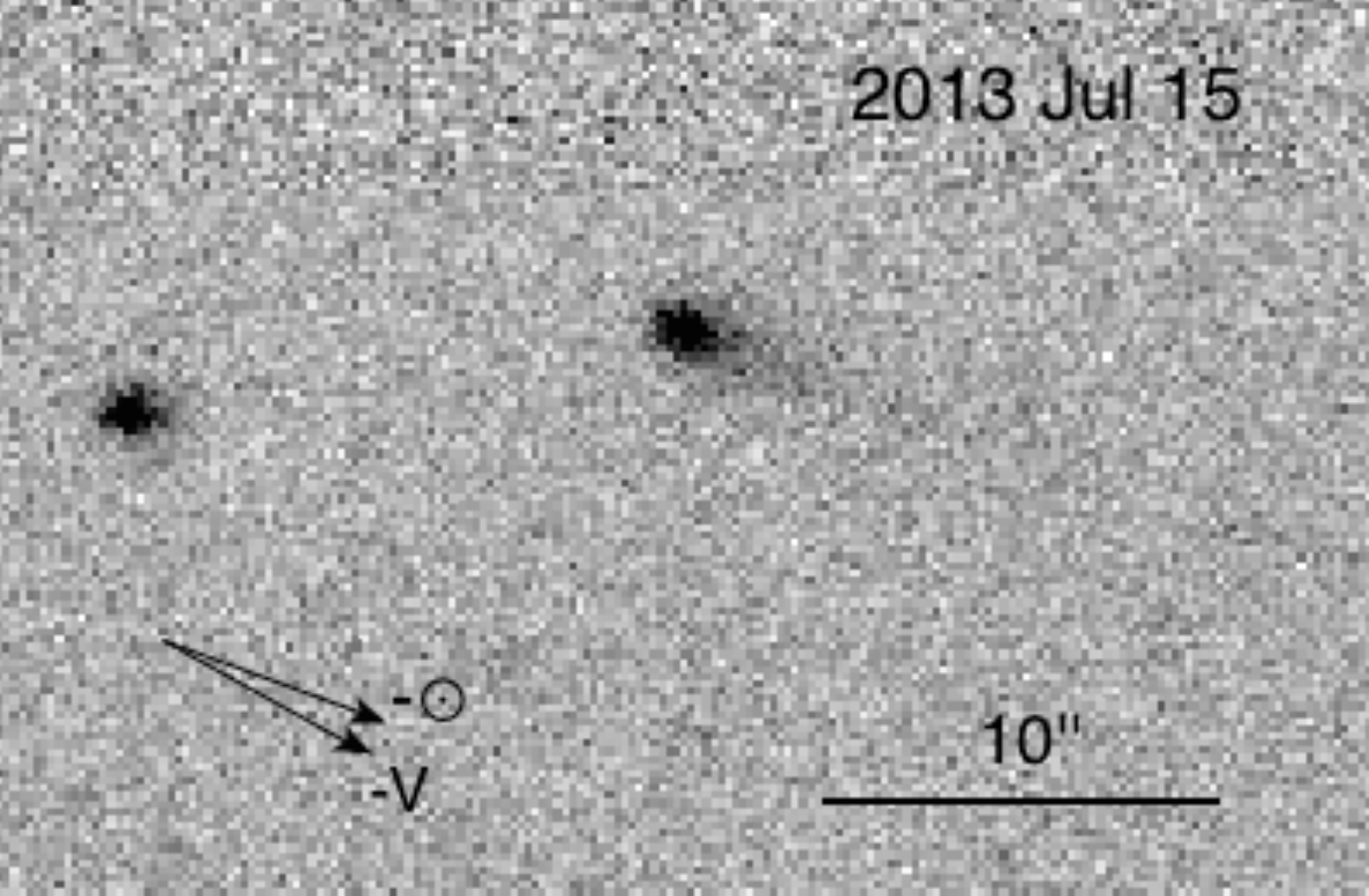}
\caption{Image of P3 recorded in archival CFHT/MegaCam data from UT 2013 July 15. This is a 120 s integration with the R filter. 
A scale bar, the projected anti-solar direction ($-\odot$) and the negative heliocentric velocity vector ($-V$) are indicated. Celestial north points up and east points to the left.
\label{2013}}
\end{figure}

\begin{figure}
\centering
\includegraphics[width=1.0\columnwidth]{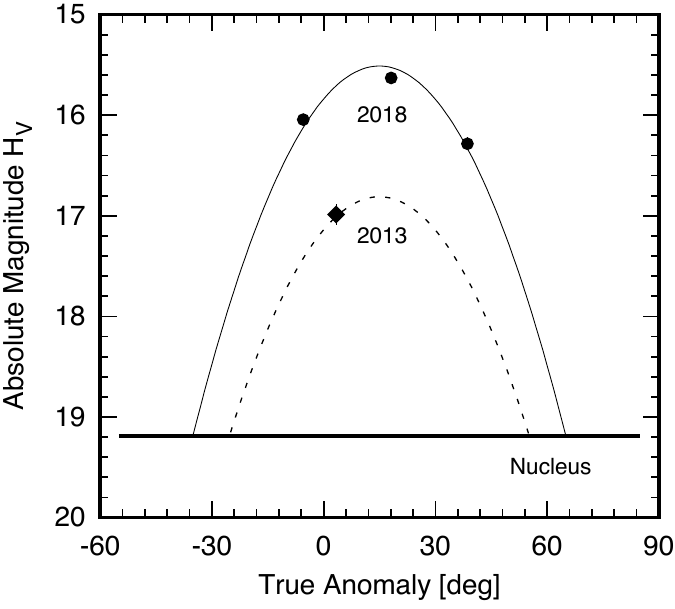}
\caption{Absolute magnitudes as a function of the true anomaly, showing the difference in activity between the 2013 and 2018 perihelion passages. Solid and dashed lines mark the best-fit quadratic functions and are added to guide the eye.
\label{comp}}
\end{figure}

\subsection{Ejection Mechanism}

The presence of a persistent anti-solar tail at each epoch of observation (Figure \ref{images}) can be naturally explained by sublimation.
More convincing evidence of a sublimation origin for the activity is the existence of archival data showing the recurrence of activity. We identified and analyzed archival data from the Canada-France-Hawaii 3.6~m telescope (CFHT) obtained with MegaCam on UT 2013 July 15 using an $R$ filter.
The CFHT data (Figure \ref{2013}) show a prominent dust tail when at heliocentric distance $R$ = 1.794 AU (perihelion was at 1.793 AU on UT 2013 July 08). This places P3 in the category of an ice-bearing MBC (Hsieh \& Jewitt 2006).

We measured the total apparent $R$ magnitude of P3 using the MegaCam calibrations but checked the result using field stars from the Pan-STARRS1.
Within a $\theta$ = 6.0$\arcsec$ radius aperture (linear radius of 8,000 km at the comet), we obtained $R$ = 20.5$\pm$0.02. Measurements were converted to $V$ using the color of the Sun $V - R$ = 0.35 (Holmberg et al. 2006). We find $V$ = 20.9 $\pm$ 0.1, where the quoted uncertainty reflects the fact that the color of P3 is not known.
The corresponding absolute $V$ magnitude computed from Equation (1) is $H_V$ = 17.0 $\pm$ 0.1, about $\sim$1 mag fainter than the 8,000~km aperture measurement from 2018 (Table 2), albeit with considerable uncertainty.

Figure \ref{comp} compares total absolute magnitudes from P3's 2013 and 2018 active periods as a function of the true anomaly, $\nu$.
The figure shows that the absolute magnitude in 2013  ($H_V \sim$ 17.0) is fainter than the extrapolated absolute magnitude in 2018  (when P3 had a similar true anomaly of $\nu \sim 0\degr$) by 1.2 mag, corresponding to a factor of $\sim$3.
This indicates that activity in P3 has increased from orbit to orbit, presumably as a result of the progressive exposure of a larger (but still small) area of sublimating ice.
Possible causes are exposure of new ice by slope collapse, edge erosion of the sublimating ice patch, or surface movement caused by rotational instability.

\begin{figure}
\centering
\includegraphics[width=1.0\columnwidth]{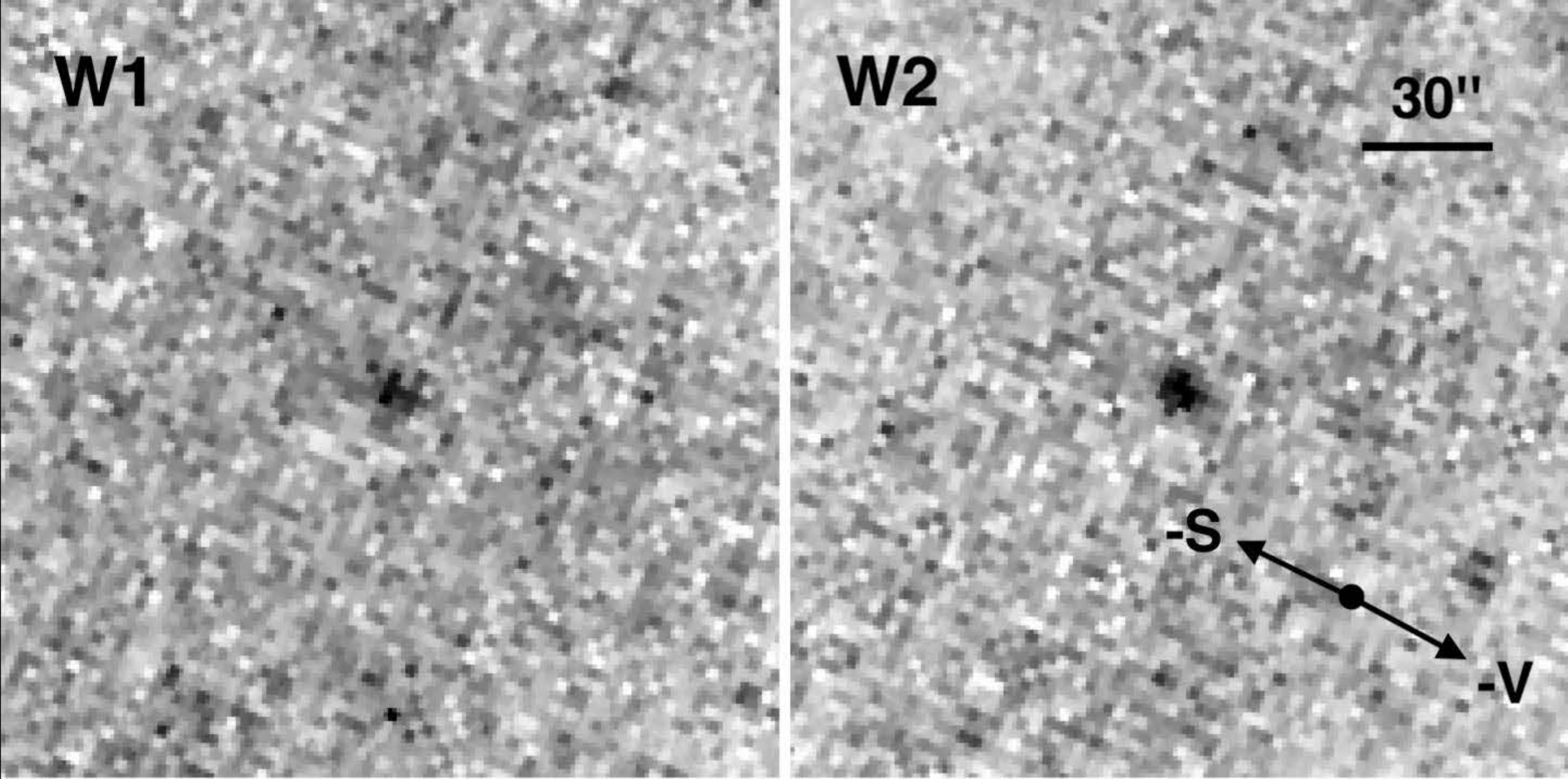}
\caption{Stacked composite images of P3 from archival NEOWISE data taken on UT 2018 December 1--2, in the 3.4 $\mu$m (``W1'', left) and 4.6 $\mu$m  (``W2'', right) channels.
A scale bar, the projected anti-solar direction ($-S$) and the negative heliocentric velocity vector ($-V$) are indicated. Celestial north points up and east points to the left.
\label{wise}}
\end{figure}

\begin{figure}
\centering
\includegraphics[width=1.0\columnwidth]{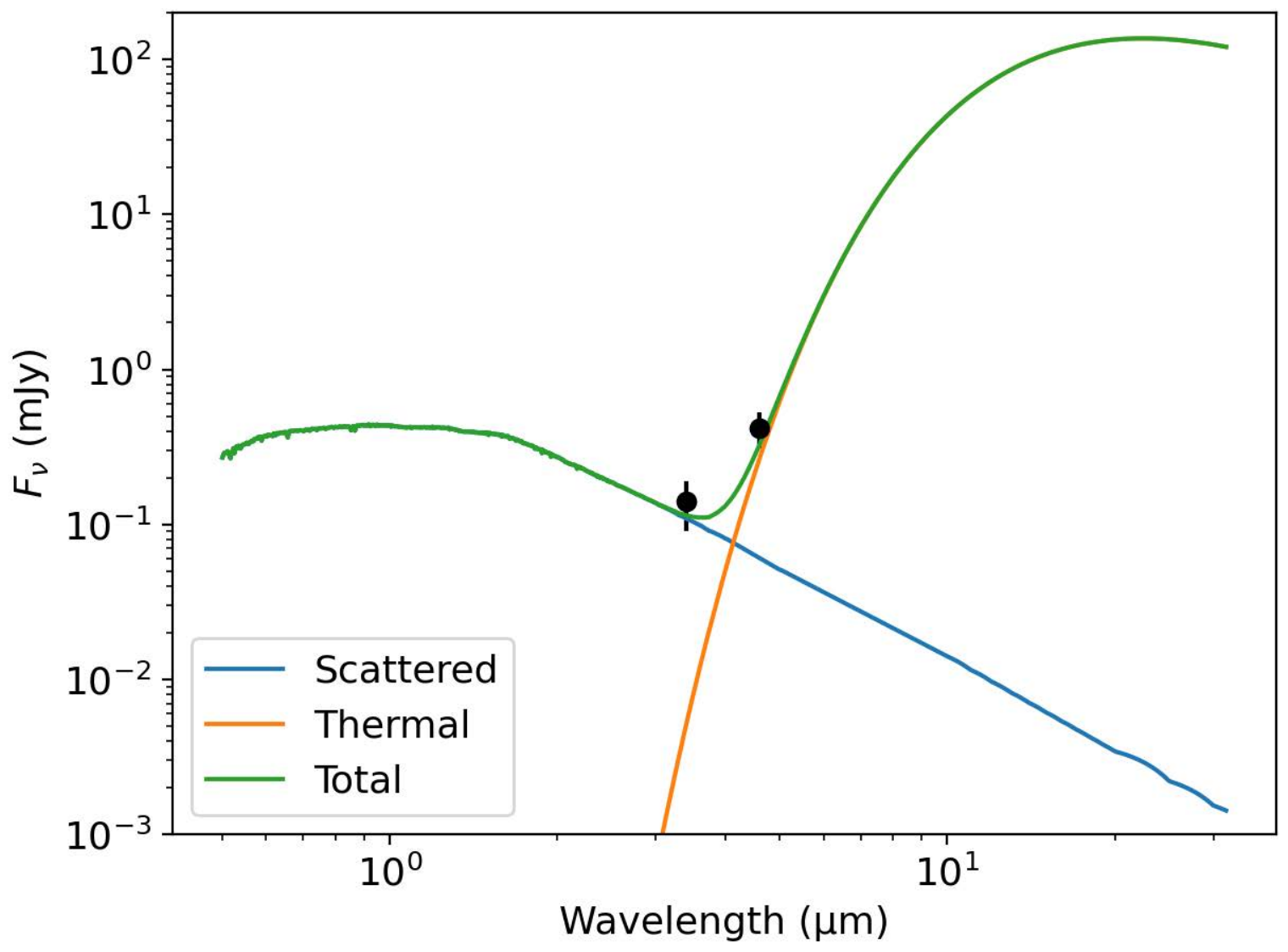}
\caption{The scattered sunlight (blue curve) and thermal emission (orange curve) fit to the 2018 Dec 1 NEOWISE data.  The black circles show measurements from the NEOWISE data, consistent with the combined reflected plus thermal contributions (green curve) from coma dust. No 4.6~$\mu$m band excess is detected.
\label{spec}}
\end{figure}

We additionally identified and analyzed archival data from the NEOWISE (Mainzer et al. 2011, 2014) obtained on UT 2018 December 1--2 in the W1 (3.4 $\mu$m) and W2 (4.6 $\mu$m) channels.
For each filter, 16 images were stacked using the online WISE MOS Search (Figure \ref{wise}).
Within a $\theta$ = 8.25$\arcsec$ radius aperture, we obtained flux densities  in W1 ((0.14$\pm$0.05) mJy) and W2 ((0.42$\pm$0.11) mJy), respectively. 
In Figure \ref{spec}, we show the W1 and W2 flux density measurements together with blue and orange curves to indicate the spectrum of scattered sunlight and the thermal emission from P3, respectively.  The scattered sunlight spectrum has been normalized to fit the W1 measurement.  The thermal emission spectrum was calculated using the total cross-section indicated by W1, assuming a dust temperature 227~K (10\% warmer than the isothermal blackbody temperature) and an emissivity to albedo ratio $\epsilon / A = 3.5$ (Kelley et al. 2016; see Equations (1)--(3) and references therein).   The green curve in Figure \ref{spec} shows the sum of the scattered and thermal emission curves.  Evidently, the composite spectrum is consistent with the data and provides no evidence for an excess in W2 that might have been associated with gaseous (CO or CO$_2$) emission (c.f.~Bauer et al. 2015). 
Similarly,  null detections of gas production have been reported in other active asteroids (Bauer et al. 2012; Snodgrass et al. 2017).

\subsection{Particle Properties}

To explore the properties of the ejected dust, we conducted a series of simulations of dust dynamics taking into account both solar gravity and radiation pressure. We created model images of P3 using a Monte Carlo dynamical procedure developed in Ishiguro et al. (2007).
As a starting point, we used the model parameters derived by Hsieh et al. (2009).
They assumed the terminal ejection velocity $V = V_0 \beta^{u_1} r_H^{-u_2}$ for $10^{-4} \le \beta \le 10^{-2}$, with $V_0$ = 25 m s$^{-1}$, $u_1$ = 0.5, $u_2$ = 0.5 and $r_H$ expressed in AU.
The model assumes that dust particles are ejected continuously, in a sunward cone with half-opening angle $\omega = 45\degr$, and follow a differential power-law size distribution with index $q = -3.5$ in the $\beta$ range of $\beta_{\rm min} \leq \beta \leq \beta_{\rm max}$.
In the new model for P3, the parameters $V_0$, $\beta_{\rm min}$, $\beta_{\rm max}$, and the onset time of dust ejection, $t_0$, were treated as variables, and the remaining parameters were used the same as in Hsieh et al. (2009).

\begin{figure*}
\centering
\includegraphics[width=0.9\textwidth]{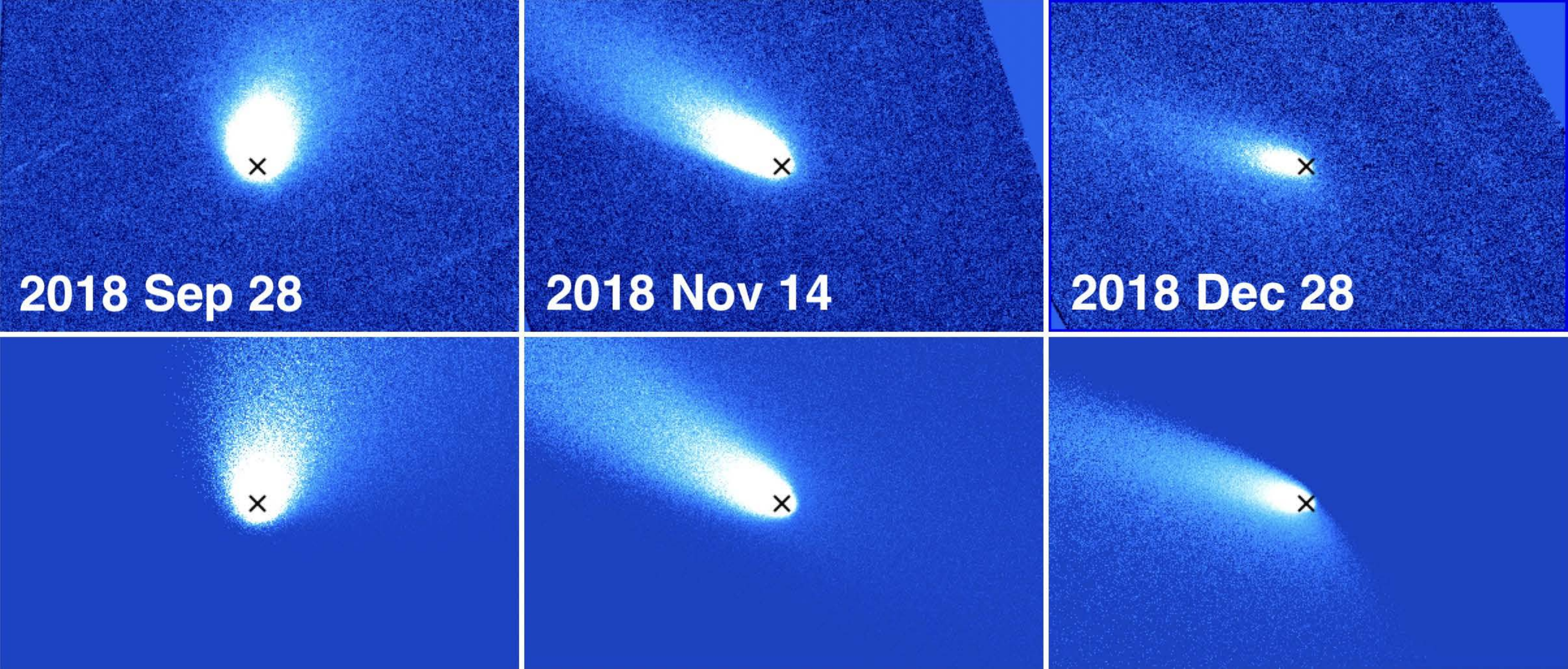}
\caption{Comparison between HST images (top) and Monte Carlo models (bottom) at three epochs of observation.
Adopted model parameters are described in the text.
Each panel shows a region $30\arcsec \times 25\arcsec$. The nucleus location is marked with a black cross in each panel. Celestial north points up and east points to the left.
\label{model}}
\end{figure*}

We created a number of model images using different parameters and then visually compared them to the observations.
With small modulation of the parameters, we found plausible solutions that could reproduce the direction, extent and overall shape of the coma.
Figure \ref{model} compares the observations with the models.
Dust ejection is assumed to begin in 2018 July (3 months before perihelion) to match the coma direction.
We find the minimum and maximum $\beta$ values between $1\times10^{-4} \le \beta_{\rm min} \le 2\times10^{-4}$ and $1\times10^{-2} \le \beta_{\rm max} \le 5\times10^{-2}$, respectively, where $V_0 = (40\pm10)$ m s$^{-1}$.
This corresponds to the dust velocity $V = 40 \sqrt{\beta/1.78} = 30\sqrt{\beta}$ m s$^{-1}$, broadly consistent with $V_{\perp} \sim 20\sqrt{\beta}$ m s$^{-1}$ inferred from the perpendicular profile (Section \ref{perpendicular}).
Similar model parameters were found for MBCs 238P, 324P, 358P, and P/2017 S5 (Hsieh et al. 2009; Moreno et al. 2011, 2013; Jewitt et al.~2019), indicating that these objects share similar properties of the ejected dust.

The best-fit parameters from the model indicate that the minimum and maximum particle radii are $a_{\rm min} \sim$ 20--100  $\mu$m and $a_{\rm max} \sim$ 0.5--1 cm, respectively.  We take $a_{\rm min}$ = 50 $\mu$m and $a_{\rm max}$ = 0.5 cm, yielding $\overline{a} = (a_{\rm min} a_{\rm max})^{1/2}$ = 0.5 mm.   We estimate an order of magnitude mass loss using

\begin{equation}
M_d = \frac{4}{3} \rho \overline{a} \Delta C_e
\end{equation}

\noindent where $\rho$ = 1000 kg m$^{-3}$ is the assumed particle density, $\overline{a}$ is the mean particle radius and $\Delta C_e$ is the change in the scattering cross-section (where brightening was observed).  We take $\overline{a}$ = 0.5 mm,  $\Delta C_e$~=~(11.2$\pm$0.7) km$^2$ between September 28 and November 14 (Table \ref{phot}) to obtain a mass ejection $M_d \sim 7.5\times10^6$ kg.  If ejected steadily over this 47 day interval, the average dust ejection rate would be $dM_d/dt \sim$ 2 kg s$^{-1}$.  
We note that this is a lower limit to the average mass loss rate, as dust moving out of the aperture was not considered.

We solved the energy balance equation for an exposed water ice surface sublimating in equilibrium with sunlight. At $r_H$ = 1.78 AU ($T$ = 196 K), we find that ice would sublimate at the specific rate $F_s$ = 1.1$\times$10$^{-4}$ kg m$^{-2}$ s$^{-1}$.  The area of exposed ice needed to supply dust at a rate, $dM_d/dt \sim$ 2 kg s$^{-1}$, is given by 

\begin{equation}
A_s = \frac{dM_d/dt}{f_{dg} F_s}
\label{subl_area}
\end{equation}

\noindent where $f_{dg}$ is the ratio of the dust to gas production rates.  We adopt $f_{dg}$ = 10 (Fulle et al. 2016; Reach et al. 2000) to find $A_s$ = 1800 m$^2$ ($\sim$0.03\% of the surface of a spherical nucleus of radius 700 m) corresponding to a circular, sublimating patch as small as $r_s = (A_s/\pi)^{1/2}$ $\sim$24 m in radius.

\begin{figure}
\centering
\includegraphics[width=1.0\columnwidth]{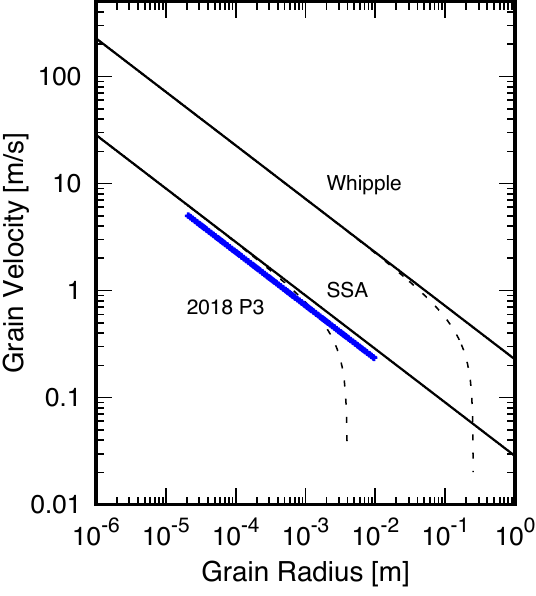}
\caption{The blue line shows the measured dust grain ejection velocity from P3 as a function of grain radius.  Black lines show the size-velocity relation from the Small Source Approximation (SSA) and the Whipple model, respectively. The dashed lines show the full solutions including nuclear gravity.
\label{whipple}}
\end{figure}

In Figure \ref{whipple}, the empirical dust grain ejection velocity from P3 (obtained from the Monte Carlo model) is compared with the velocities predicted using the classical Whipple model (Whipple 1951) and the small source approximation (SSA) model (Jewitt et al. 2014). 
In the SSA, a small source (length scale $r_s \sim$24 m) limits the acceleration length for gas-entrained dust particles and so leads to lower dust velocities.
The empirical dust speeds (blue line) are an order of magnitude smaller than predicted by the Whipple model but consistent with those of the SSA model for a vent of the measured size.
Ejection velocities of 5 mm particles are comparable to the 0.3 m s$^{-1}$ gravitational escape speed of the (non-rotating) nucleus, while smaller particles are launched at speeds faster than the gravitational escape velocity.

\subsection{Dynamics}

P3 has a high eccentricity ($e$ = 0.415), perihelion distance ($q$ = 1.756 AU) close to the aphelion of Mars ($Q_{\rm Mars}$ = 1.666 AU), and $T_J$ value ($T_J = 3.096$) in the dynamical boundary region between asteroids and comets.
Furthermore, the semimajor axis ($a$ = 3.007 AU) lies close to the 9:4 mean-motion resonance with Jupiter at $a_{9:4}$ = 3.029 AU, suggesting that this resonance may induce dynamical instability.

\begin{figure}
\centering
\includegraphics[width=1.0\columnwidth]{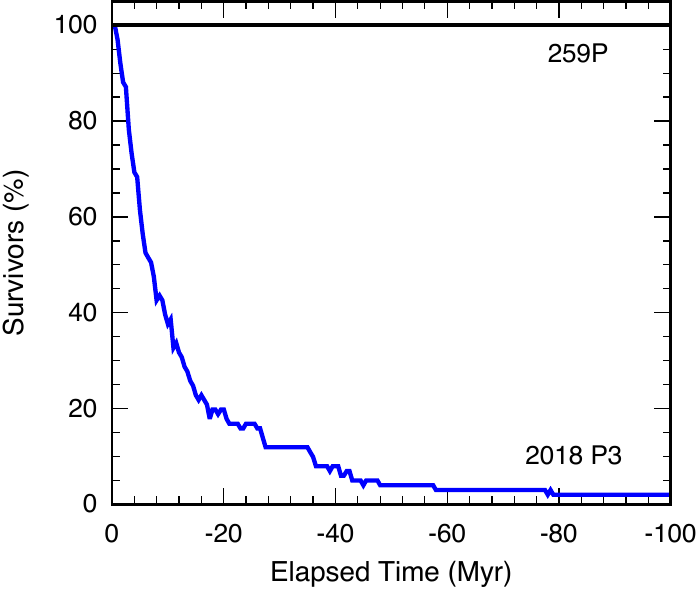}
\caption{Percentage of P3 and 259P clones that remain in main-belt-like orbits ($2.064<a<3.277$ AU and $q>1.65$~AU and $Q<4.50$~AU) in the backward dynamical evolution.
P3 survivors decrease exponentially on an e-folding time $\sim$12 Myr.
During the whole computed period of 100~Myr, 98\% of the P3 clones left the main belt, while all 259P clones remained in the main belt.
\label{clones}}
\end{figure}

To assess whether P3 is native to its current location, we investigated its long-term dynamical stability.
We generated 1000 Gaussian distributed clones in orbital element space, centered on the osculating orbital elements of P3 and used $\sigma$ values equal to the orbital element uncertainties.
The osculating orbital elements and their 1$\sigma$ uncertainties were retrieved from the JPL database at epoch 2018 October 21.
A more realistic clone can be obtained from a multivariate normal distribution with an orbital covariance matrix, but it is not used here for the present purpose.
We integrated the orbits backward for 100 Myr using the Mercury N-body integration package (Chambers 1999), where we set the non-gravitational force equal to zero.
For comparison, we additionally generated and integrated 1000 clones of 259P, an MBC that is close to the 8:3 mean-motion resonance with Jupiter (Jewitt et al. 2009). These clones were generated with small deviations to reflect the small orbital element uncertainties of 259P.
Figure \ref{clones} shows the percentage of P3 and 259P clones that remain in main-belt-like orbits ($2.064<a<3.277$ AU and $q>1.65$~AU and $Q<4.50$~AU; following Hsieh \& Haghighipour 2016) in the backward dynamical evolution.
The number of P3 survivors decreases exponentially on an e-folding time $\sim$12 Myr.
During the whole computed period of 100~Myr, 98\% of the P3 clones left the main belt (implying that the orbits are highly unstable), while all 259P clones remained in the main belt.
The orbit of P3 is chaotic and, therefore, not possible to follow back except for a short period of time.  Below we consider two possible sources for P3.

\begin{table*}
\caption{Orbital Elements of P/2018 P3 and 233P
\label{233P}}
\centering
\begin{tabular}{lccccccccc}
\hline\hline
Object & $a$\tablefootmark{a} & $e$\tablefootmark{b}& $i$\tablefootmark{c}   & $T_J$\tablefootmark{d} & $q$\tablefootmark{e} & $Q$\tablefootmark{e}  & $r_e$\tablefootmark{f} & $Q_{CO_2}$\tablefootmark{g} & Ref.\tablefootmark{h}
\\
& (AU) & & (deg) & & (AU) & (AU) & (km) & (mol~s$^{-1}$) \\
\hline
P/2018 P3 & 3.007 & 0.415 & 8.90 & 3.096 & 1.756 & 4.257 & $<$0.71 & -- & [1]  \\
233P & 3.033 & 0.410 & 11.27 & 3.081 & 1.787 & 4.279 & $\sim$0.54 & 1.1$\times10^{25}$ & [2] \\
\hline
\end{tabular}
\tablefoot{
\tablefoottext{a}{Semimajor axis in AU.}
\tablefoottext{b}{Orbital eccentricity.}
\tablefoottext{c}{Orbital inclination (degrees).}
\tablefoottext{d}{Tisserand parameter with respect to Jupiter.}
\tablefoottext{e}{Perihelion, $q$ and aphelion, $Q$, distances, in AU.}
\tablefoottext{f}{Effective circular radius, in km.}
\tablefoottext{g}{CO$_2$ production rate at $r_H$ = 1.8 AU, in molecules~s$^{-1}$.}
\tablefoottext{h}{References: [1] This work; [2] Bauer et al. (2015).}}
\end{table*}

\begin{figure*}
\centering
\includegraphics[width=0.9\textwidth]{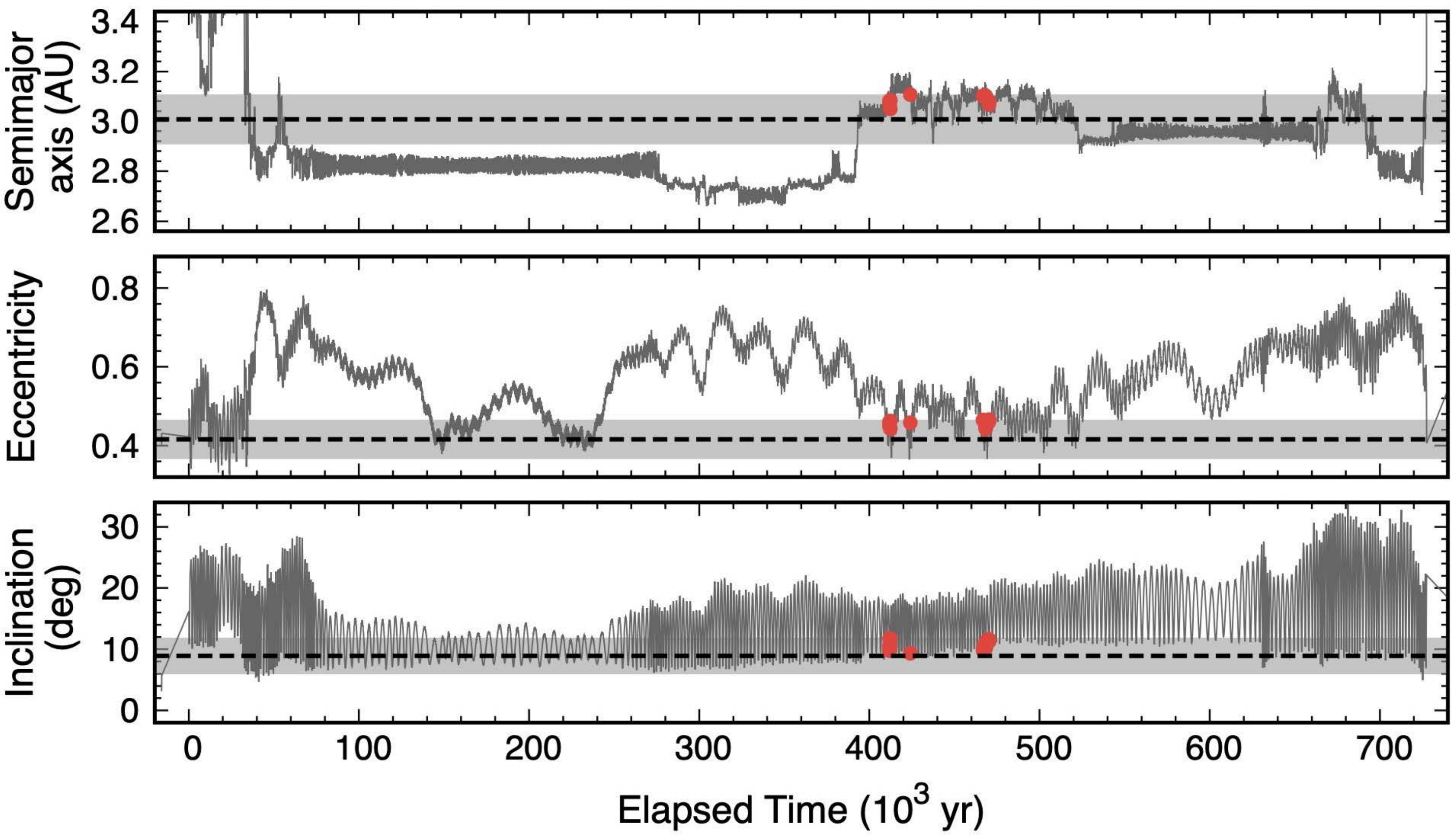}
\caption{Semimajor axis (top), eccentricity (middle), and inclination (bottom) as a function of time for a synthetic JFC from Brasser \& Morbidelli (2013). The shaded regions indicate where orbital elements are similar to those of P3, where $a=a_{\rm P3} \pm 0.1$~AU, $e = e_{\rm P3} \pm 0.05$, and $i = i_{\rm P3} \pm 3\degr$.  The points where the orbital elements of the synthetic comet meet all these criteria are marked in red.
\label{JFC}}
\end{figure*}

Table \ref{233P} lists the orbital elements of P3 and 233P, which are remarkably similar and suggest the same origin.
Hsieh et al. (2018) used synthetic Jupiter-family comets (JFCs) studied by Brasser \& Morbidelli (2013) to find a dynamical path from the Kuiper Belt to the orbit of  233P (Figure 29 of Hsieh et al. 2018).
Because of the similarity of their orbital elements, we find that a single synthetic JFC could take on both 233P-like and P3-like orbital elements at some point during their evolution (Figure \ref{JFC}).
We infer that there is a small but  non-zero probability that JFCs can evolve into P3-like orbits.
On the other hand, inspection of synthetic JFCs did not find a dynamical path from the Kuiper Belt to the orbit of 259P (R. Brasser 2018, private communication), consistent with the conclusion in Jewitt et al. (2009) that it originated elsewhere in the main belt.

Lastly, we consider the possibility that P3 originated in the main belt. The dynamical evolution of main belt objects into mean-motion resonances due to  Yarkovsky drift, and subsequent increase of eccentricity and inclination is a possible scenario (e.g. Hsieh et al. 2020; Kim et al. 2014). Since our integrations did not include non-gravitational forces such as the Yarkovsky effect, we leave their role open in the dynamical evolution of P3.

\section{SUMMARY}

We present Hubble Space Telescope measurements of active asteroid P/2018 P3 taken on three occasions between UT 2018 September 28 ($r_H$ = 1.758 AU, inbound) and 2018 December 28 ($r_H$~=~1.877 AU, outbound).  
We additionally identify and analyze archival CFHT data (showing P3 to have been active in 2013) and NEOWISE data (showing the absence of 4.6~$\mu$m band excess).  We find that

\begin{enumerate}

\item Sublimation explains the protracted nature of the activity, and its recurrence near perihelion in both 2013 and 2018.
The null detection of a companion or fragments or multiple tails (in high-quality HST data) also rules out the other sources of activity (impact or spin-up).

\item The absence of 4.6~$\mu$m band excess indicates zero or negligible CO or CO$_2$ gas production.

\item Photometry sets a limit to the effective radius of the nucleus at $r_e <$ 0.7 km (geometric albedo $p_V$ = 0.04 assumed).

\item Properties of the ejected dust are remarkably consistent with previously studied MBCs (continuous emission of $\sim$0.05--5 mm particles at 0.3--3 m s$^{-1}$ speeds).
This suggests that both P3 and MBCs have small active areas, as they physically age in the main belt.
The average dust production rate, $dM_d/dt \gtrsim$ 2 kg s$^{-1}$, could be supplied by the sublimation of water ice covering as little as $A_s \sim$1800 m$^2$.

\item Our dynamical analysis suggests that the orbit of P3 is unstable on timescales $\sim$10 Myr and therefore that P3 originated elsewhere.
98\% of the P3 clones left the main belt for the whole backwards computed period of 100~Myr.

\end{enumerate}

We speculate that P3 has recently arrived in the main belt from a source region (either the Kuiper Belt or elsewhere in the main belt) and has been physically aged at its current location ($q \sim$ 1.76 AU), finally becoming indistinguishable from a weakly sublimating MBC in terms of its dust properties.
Whatever the source of P3, given the dynamical instability of its current orbit, P3 should not be used to trace the native distribution of main-belt ice.

\bigskip
\begin{acknowledgements}
We thank the anonymous referee for comments on the manuscript and Ramon Brasser for providing the orbital elements of synthetic JFCs. Based on observations made under GO 15623 with the NASA/ESA Hubble Space Telescope obtained from the Space Telescope Science Institute,  operated by the Association of Universities for Research in Astronomy, Inc., under NASA contract NAS 5-26555.
This publication makes use of data products from the Near-Earth Object Wide-field Infrared Survey Explorer (NEOWISE), which is a joint project of the Jet Propulsion Laboratory/California Institute of Technology and the University of Arizona.

Y.K. and J.A. acknowledge funding by the Volkswagen Foundation. J.A.'s contribution was made in the framework of a project funded by the European Union's Horizon 2020 research
and innovation programme under grant agreement No 757390 CAstRA.
\end{acknowledgements}

{\it Facilities:}  HST (WFC3), NEOWISE.

{\it Software:} sbpy (Mommert et al. 2019).

\end{document}